# Plasmon Collapse in a Directional Strange Metal

Jiade Li[1,2#*], Peiyi He[1,2#], Deng Hu[3,4#], Jiakai Wang[1,2], Jifeng Shao[5], Wan-Shun Du[6], Xuetao Zhu[7,8], Jiandong Guo[7,8], Zhiwei Wang[3,4,5*], Peng Gao[1,2,9,10*]

[1] *International Center for Quantum Materials, School of Physics, Peking University, Beijing 100871, China*

[2] *Electron Microscopy Laboratory, School of Physics, Peking University, Beijing 100871, China*

[3] *Centre for Quantum Physics, Key Laboratory of Advanced Optoelectronic Quantum Architecture and Measurement (MOE), School of Physics, Beijing Institute of Technology, Beijing 100081, China*

[4] *Beijing Key Laboratory of Quantum Matter State Control and Ultra-Precision Measurement Technology, Beijing Institute of Technology, Beijing 100081, China*

[5] *Marine Science and Technology Domain, Beijing Institute of Technology, Zhuhai 519088, China*

[6] *International Quantum Academy, Shenzhen 518048, China*

[7] *Beijing National Laboratory for Condensed Matter Physics and Institute of Physics, Chinese Academy of Sciences, Beijing 100190, China*

[8] *School of Physical Sciences, University of Chinese Academy of Sciences, Beijing 100049, China*

[9] *Interdisciplinary Institute of Light-Element Quantum Materials and Research Center for Light-Element Advanced Materials, Peking University, Beijing 100871, China*

[10] *Tsientang Institute for Advanced Study, Zhejiang 310024, China*

[#] These authors contributed equally to this work.

[*] Correspondence address: Jiade Li (jdli@pku.edu.cn); Zhiwei Wang (zhiweiwang@bit.edu.cn); Peng Gao (pgao@pku.edu.cn).

**Strange metals are widespread in strongly correlated quantum matter and closely intertwined with high-temperature superconductivity and quantum criticality[1], yet their microscopic charge dynamics remain fundamentally unresolved. Whether well-defined collective modes survive in an incoherent electronic fluid or dissolve into overdamped continua remains intensely debated[2-9]. Here, employing high-momentum-resolution inelastic electron scattering complemented by direction-resolved transport, we show that the van der Waals compound $MoOCl_2$, a "bad" metal known to host unusually "good" plasmons[10], exhibits a striking dichotomy in electronic coherence and charge dynamics. Transport reveals coherent Fermi liquid behavior along one in-plane crystallographic axis and an incoherent strange metal response along the orthogonal axis. This dichotomy extends directly to the charge response. The Fermi liquid direction supports a well-defined, long-lived plasmon, whereas the strange metal direction exhibits an overdamped excitation that progressively loses its collective character with increasing momentum. Beyond a critical momentum ($q_c \approx 0.008$ Å$^{-1}$), the mode undergoes a reversal into negative dispersion and ultimately collapses into a broad, nearly featureless continuum with a robust high-energy boundary ($E_c \approx 1.75$ eV). Remarkably, thickness-dependent measurements show that both $q_c$ and $E_c$ remain nearly unchanged despite substantial modifications of Coulomb screening, revealing them as intrinsic scales of incoherent charge density fluctuations. These results establish a unique single-material platform in which coherent and incoherent electronic states are partitioned into in-plane orthogonal channels, and demonstrate that incoherent electronic fluids retain intrinsically structured charge dynamics beyond the quasiparticle paradigm.**

Strange metals represent one of the most enigmatic electronic states in modern condensed matter physics. Their hallmark linear-in-temperature resistivity, typically accompanied by violations of the Mott-Ioffe-Regel (MIR) limit and near-Planckian dissipation, signals a profound breakdown of Landau Fermi liquid theory[1]. In a conventional Fermi liquid, long-lived quasiparticles not only constitute the low-energy single-particle spectrum but also provide a unified microscopic basis for collective charge behavior[11]. In strange metals, however, the breakdown of long-lived quasiparticles leaves the electronic fluid intrinsically incoherent, rendering its charge dynamics fundamentally elusive.

The dynamical charge density response provides a direct probe of this problem, revealing how charge density fluctuations correlate, propagate, and dissipate without invoking a quasiparticle description[12,13]. Yet existing experiments have yielded sharply conflicting pictures of the strange metal charge response. Some have observed well-defined plasmons[2,3,10], suggesting that electronic incoherence does not necessarily preclude coherent collective charge motion. Others have detected strongly overdamped excitations or featureless continua[5-7], pointing to a severe suppression of coherent collective behavior. This discrepancy has locked the field into a long-standing binary debate over whether plasmons survive or perish. However, the presence or absence of a collective mode captures only one facet of the broader charge response. A microscopic understanding of strange metal charge dynamics requires mapping the full dynamical charge response across energy-momentum space, tracing its evolution and identifying its underlying physical scales.

Nevertheless, progress has long been limited by a critical experimental gap. Conventional optical spectroscopy[14] is restricted to the light cone, allowing only small momentum transfers. Near-field[15] and nanostructure-assisted optical[16] approaches can access finite momenta but remain primarily sensitive to coherent collective modes, leaving strongly damped excitations and incoherent continua difficult to resolve. Inelastic scattering techniques, such as electron energy loss spectroscopy[17,18] and resonant inelastic X-ray scattering[19], directly probe finite-momentum charge excitations, but their limited momentum resolution hinders access the crucial long-wavelength regime. As a result, a full energy-momentum landscape of strange metal charge dynamics has remained experimentally inaccessible.

Here, using defocus-engineered scanning transmission electron microscopy-electron energy loss spectroscopy (STEM-EELS) with momentum resolution down to 0.001 $Å^{-1}$, combined with directional transport, we directly map and contrast coherent versus incoherent charge dynamics in $MoOCl_2$. We reveal an orthogonal electronic dichotomy, where a coherent Fermi liquid and an incoherent strange metal response coexist along perpendicular in-plane directions. Correspondingly, long-lived plasmons propagate along the Fermi liquid direction, whereas charge excitations along the strange metal direction rapidly lose collective character and collapse into a broad continuum. We identify two characteristic scales of the strange metal charge response: a momentum $q_c \approx 0.008$ $Å^{-1}$ initiating spectral reconstruction and a high-energy boundary $E_c \approx 1.75$ eV delimiting the incoherent continuum. Remarkably, both scales remain unchanged across substantial thickness-induced variations in Coulomb screening, identifying them as intrinsic features of the incoherent charge dynamics. By tracing the full evolution of charge excitations across energy-momentum space, we reveal how collective charge motion loses coherence and establish the intrinsic structure of strange metal charge dynamics beyond the quasiparticle paradigm.

## Orthogonal electronic dichotomy in $MoOCl_2$

$MoOCl_2$ is a layered van der Waals metal with a rectangular in-plane crystal lattice (Fig. 1a). Its orthogonal crystallographic directions are distinguished by the Mo-O chains (a-axis) and the Peierls-distorted Mo-Cl framework (b-axis), giving rise to pronounced electronic anisotropy[20]. Previous studies identified $MoOCl_2$ as a bad metal, with the sheet resistance exceeding the resistance quantum, whereas near-field optical measurements observed remarkably long-lived propagating plasmon polaritons[10,21]. This apparent paradox of "good" plasmons in a "bad" metal motivates a direction-resolved examination of its electronic behavior.

We therefore measured the resistivity along the two in-plane orthogonal crystallographic axes on high-quality single crystals, as confirmed by the sharp (00*l*) X-ray diffraction peaks (Fig. 1b). Along the b-axis, the resistivity $\rho$ is anomalously large and exhibits two distinct regimes (Fig. 1c). Below approximately 110 K, it follows a nearly quadratic temperature dependence, $\rho \sim T^{1.96}$, whereas above this temperature it crosses gradually into an extended $T$-linear regime, $\rho \sim T^{0.98}$. Notably, the $T$-linear resistivity exceeds the estimated MIR limit of $521\ \mu\Omega \cdot \mathrm{cm}$ (Methods) and continues without any sign of saturation up to the highest temperature measured. The persistent $T$-linear resistivity and MIR violation are classic hallmarks of strange metal behavior, where a conventional coherent quasiparticle description breaks down.

In contrast, the orthogonal a-axis behaves fundamentally differently, with its resistivity being only about 0.05 times that along the b-axis throughout the measured temperature range (Fig. 1d). Over the entire measured temperature range, the resistivity remains well described by $\rho \sim T^{2.08}$ and stays far below the corresponding MIR limit ($978\ \mu\Omega \cdot \mathrm{cm}$), establishing conventional Fermi liquid transport. $MoOCl_2$ therefore realizes an unconventional in-plane electronic dichotomy, with Fermi liquid and strange metal response encoded along two orthogonal crystallographic directions.

Such an orthogonal electronic dichotomy resembles the behavior predicted in holographic models of anisotropic quantum critical continua[22], but has not, to our knowledge, been established experimentally. Crucially, this behavior fundamentally departs from familiar electronic dichotomies in typical correlated systems. In cuprates, electronic dichotomy predominantly separates distinct sectors of momentum space[23,24], whereas in iron-based superconductors it is primarily orbital selective[25,26]; in either case, these sectors remain intermixed in macroscopic transport. In $MoOCl_2$, by contrast, electronic coherence itself is encoded directly into crystallographic directions in real space, with coherent and incoherent regimes disentangled into orthogonal transport channels. This directional separation provides a unique platform to directly compare coherent and incoherent charge dynamics within a single material.

## Directional propagation of plasmon polaritons

The orthogonal electronic dichotomy raises a direct question: are the previously observed long-lived plasmon polaritons supported by the incoherent electronic fluid itself, or do they arise selectively from the coherent Fermi liquid channel? To address this, we employ STEM-EELS with a conventional large-convergent-beam configuration (Methods) to directly visualize the propagation of collective excitations along the two orthogonal directions. A focused electron probe positioned near the edge of an 8-layer flake excites a broadband electromagnetic response[27], with propagating plasmon polaritons manifesting as distance-dependent modulations in the energy-loss intensity via interference with edge-reflected waves (Fig. 2a, b). Electron diffraction identifies the reciprocal a*- and b*-axes and thereby determines the corresponding real-space a- and b-oriented flake edges (Fig. 2c).

Along the a-axis, the energy-loss map displays pronounced, regularly evolving interference fringes extending from the edge (white dashed line) into the flake interior (Fig. 2d). The corresponding energy distribution curves (EDCs) exhibit systematic shifts of the polaritonic peaks with distance (Fig. 2f), reconstructing a well-defined dispersing branch in energy-momentum space (Fig. 2h; Methods). These results establish coherent plasmon polariton propagation along the Fermi liquid direction. By contrast, along the strange metal b-axis, no comparable interference fringes, systematic peak evolution or resolved polaritonic branch is observed (Fig. 2e, g, i). The directional plasmonic response therefore resolves the apparent paradox of "good" plasmons in a "bad" metal: propagating plasmon polaritons survive only along the Fermi liquid channel but are quenched along the orthogonal strange metal channel.

Notably, a broad continuum is present along both directions and remains insensitive to probe position (Fig. 2f, g), indicating intrinsic electronic excitations rather than propagating polaritonic modes. However, the real-space scanning signal here is dominated by collective excitations with sufficient coherence and propagation length to generate interference, and therefore cannot resolve the spectral character of this non-propagating continuum. We thus turn to momentum-resolved STEM-EELS to directly track the intrinsic charge excitations in energy-momentum space along both directions.

**Collapse of coherent collective charge excitations**

Momentum-resolved STEM-EELS probes intrinsic charge excitations through the energy- and momentum-dependent loss function, providing access to the dynamical charge density response (Methods). It can therefore capture not only well-defined collective modes, but also strongly damped excitations and incoherent continua that are inaccessible to real-space propagation imaging. To resolve the complete charge response in the long-wavelength regime, we employ a recently developed defocus-engineered STEM-EELS approach based on the pioneering work of Midgley[28] (Fig. 3a, see also Methods).

The resulting energy-momentum maps reveal a pronounced directional contrast. Along the X-Γ-X (Fermi liquid) direction, a sharp plasmon branch disperses continuously to higher energy with increasing momentum (Fig. 3b). The corresponding momentum-dependent EDCs exhibit a prominent, well-resolved peak that remains well defined throughout the measured momentum range (Fig. 3d). Its narrow linewidth and regular positive dispersion demonstrate that the collective excitation retains a coherent, long-lived character along the Fermi liquid direction, consistent with the long-range plasmon polariton propagation observed in real space.

By contrast, along the Y-Γ-Y (strange metal) direction, the charge excitation exhibits severe broadening even in the long-wavelength regime (Fig. 3c, e). At small momenta, a residual plasmon-like feature remains discernible and retains a positive dispersion similar to that along the Fermi liquid direction. Despite being overdamped, with a linewidth comparable to its excitation energy (Fig. 3f and Extended

Data Fig. 5), this positive dispersion shows that the mode still retains a spatially correlated collective character. With increasing momentum, however, this coherence is rapidly lost. The mode reaches a maximum energy at a characteristic momentum $q_c \approx 0.008$ Å$^{-1}$, beyond which the dispersion bends back and softens, accompanied by a rapid loss of pole-like character (Fig. 3c, e, f). This spectral reconstruction signals a rapid loss of coherent collective spectral weight in favor of incoherent excitations. At still larger momenta, the excitation ultimately collapses into a broad, nearly featureless continuum (Fig. 3g). The nearly momentum-independent, featureless continuum signals essentially incoherent charge fluctuations lacking long-range spatial correlations. This dramatic evolution lies beyond the quasiparticle-based random phase approximation (RPA) picture, which predicts dispersive plasmons persisting throughout the relevant momentum range in $MoOCl_2$ (ref.[29]). Instead, our measurements directly capture how coherent charge excitations collapse into an incoherent continuum in the strange metal state.

The characteristic momentum $q_c$ marks the onset at which the collective charge response loses coherent character and evolves into an incoherent continuum. Its inverse defines a characteristic length, $\xi_c \sim 1/q_c \approx 12.5$ nm. On spatial scales larger than $\xi_c$, charge fluctuations retain a finite degree of collectivity, whereas below $\xi_c$, this collectivity rapidly breaks down and the response becomes dominated by the incoherent continuum. Thus, $\xi_c$ represents the characteristic spatial scale over which collective charge motion remains sustained in the strange metal state.

**Intrinsic scales of incoherent charge dynamics**

To identify the physical factors governing the collapse of the collective mode, we performed thickness-dependent measurements to tune the long-range Coulomb screening. The energy-momentum maps obtained from 2-layer and 16-layer samples reproduce the pronounced directional dichotomy observed in the 8-layer sample, with coherent Fermi liquid and incoherent strange metal charge excitations remaining robust across thickness (Fig. 4a-d).

Quantitatively, the plasmon along the Fermi liquid direction evolves systematically with thickness. With increasing layer number, the plasmon branch shifts to higher energy and develops a steeper dispersion (Fig. 4e), as expected for a collective electronic mode governed by long-range Coulomb interactions[30-32]. At larger momenta, the dispersion progressively flattens into the slow-plasmon regime associated with interband screening within a quasiparticle framework[33,34]. Notably, several additional, weakly dispersive branches emerge at lower energies in the 16-layer sample (Fig. 4b). These modes are consistent with acoustic plasmon branches generated by interlayer coupling and reflect the characteristic multibranch collective response of a layered electron system[30,31,35,36]. Furthermore, the plasmon retains a narrow linewidth across all measured thicknesses (Fig. 4e). These features establish the well-defined and long-lived character of the Fermi liquid plasmon.

Along the strange metal direction, the overdamped long-wavelength branch also exhibits a pronounced thickness dependence, with its energy increasing systematically with layer number (Fig. 4a-d). Its similar evolution to that of the well-defined plasmon along the Fermi liquid direction (Extended Data Fig. 6) provides further evidence that the long-wavelength excitation retains a residual coherent character and remains sensitive to the long-range Coulomb screening. Notably, the characteristic back-bending occurs at nearly the same momentum, $q_c \approx 0.008$ Å$^{-1}$, across all measured thicknesses (Fig. 4f, blue spheres), even though the corresponding excitation energy varies substantially (Fig. 4f, red spheres). This suggests that $q_c$ is an intrinsic momentum scale of the strange metal dynamical charge density response.

More strikingly, with increasing momentum, the initially distinct responses from different thicknesses progressively converge into the same broad, nearly featureless continuum (Fig. 4g and Extended Data Fig. 8). This spectral convergence shows that, once the coherent collective mode has collapsed, the charge response becomes insensitive to Coulomb screening, identifying the continuum as an intrinsic feature of the strange metal charge dynamics. The incoherent continuum is further terminated at a robust high-energy boundary $E_c \approx 1.75$ eV. Within this incoherent regime, $E_c$ remains nearly independent of both momentum and thickness (Fig. 4g and Extended Data Fig. 8), indicating that it is an intrinsic energy scale of the incoherent charge response. Together, $q_c$ and $E_c$ establish characteristic momentum and energy scales for the incoherent charge dynamics.

**Discussion and outlook**

The pronounced long-wavelength overdamping observed along the strange metal direction is broadly consistent with a common physical picture emerging from holographic[8,37], marginal Fermi liquids[38,39], and strongly coupled critical-boson[38] theories. In these frameworks, an incoherent critical continuum provides an intrinsic decay channel for collective charge excitations already in the long-wavelength limit. Our measurements show, however, that this long-wavelength behavior represents only the beginning of a much more dramatic evolution. In particular, the emergence of a critical momentum $q_c$ marking the onset of this reconstruction has no obvious counterpart in these existing microscopic descriptions.

By contrast, the phenomenological jamming model[40] and a recent charge-conservation analysis[41] have anticipated a characteristic momentum and emphasized its importance for understanding strange metal charge dynamics. However, neither approach incorporate a microscopic description of strange metallicity nor accounts for the magnitude of $q_c$, leaving its physical origin and microscopic significance unexplained. Our measurements place direct experimental constraints on the nature of this scale. Despite substantial thickness-induced changes in both Coulomb screening and excitation energy, $q_c$ remains essentially invariant (Fig. 4f). The robustness of $q_c$ indicates that it is not set by the Coulomb screening. Its insensitivity to the changing excitation energy also disfavors a simple interpretation in terms of an energy-dependent nonlocal transport length. Additionally, the sharp spectral reconstruction at $q_c$ differs from the smooth crossover expected from charge conservation[41]. Together, these results reveal $q_c$ as an intrinsic momentum scale governing strange metal charge dynamics, corresponding to a characteristic real-space correlation length $\xi_c \sim 1/q_c$ of the incoherent charge fluid.

The back-bending, softening, and rapid collapse of the charge excitations above $q_c$ constitute a remarkable finite-momentum evolution of the incoherent charge response. Once the collective mode collapses, the response becomes decoupled from the thickness-dependent Coulomb screening and forms a broad, nearly featureless continuum. Yet a featureless spectrum does not imply a lack of physical structure. The continuum terminates at a sharp high-energy boundary $E_c$, confining the incoherent charge response to a well-defined energy window. The significance of this finite-energy boundary has been largely overlooked: charge-conservation sum rules[13] constrain the integrated spectral weight but do not require a finite-energy termination, while existing theories either impose a phenomenologically cut off or defer to non-universal microscopic details[8,38,39]. Moreover, we find that $E_c \approx 1.75$ eV lies well below the interband transition onset near 3 eV, as indicated by previous calculations[20] and confirmed by our own measurements (Extended Data Fig. 9). This clear separation shows that $E_c$ belongs to the incoherent charge response itself and is not directly tied to higher-energy single-particle excitations. These results reveal that the well-defined energetic structure is likely associated with the strong many-body interactions underlying the incoherent electronic fluid.

Our findings move beyond the binary question of whether a plasmon survives in a strange metal. By mapping the dynamical charge response across the complete energy-momentum landscape, we reveal how collective charge motion progressively loses coherence: from an overdamped yet dispersive excitation at long wavelengths, through back-bending and softening at $q_c$, to a nearly featureless continuum bounded by $E_c$. These results demonstrate that, even when quasiparticles break down, the incoherent charge dynamics retain an internal structure defined by intrinsic momentum and energy scales. The complete energy-momentum trajectory and its intrinsic scales establish quantitative benchmarks for understanding incoherent charge dynamics.

More broadly, our results establish $MoOCl_2$ as an unusual electronic system in which coherent and incoherent states coexist along in-plane orthogonal crystallographic directions. This intrinsic dichotomy naturally resolves the apparent "good" plasmon–"bad" metal paradox and provides a microscopic basis for the hyperbolic plasmon polaritons in $MoOCl_2$, which have attracted broad interest by combining low loss with spectral coverage extending into the visible[42-45]. These results provide a unified picture of electronic coherence spanning electrical transport, collective charge dynamics and optical response over a wide range of energy scales. Selectively accessing and controlling these orthogonal coherent and incoherent channels will open new avenues for the directional routing of charge and light within a single quantum material.

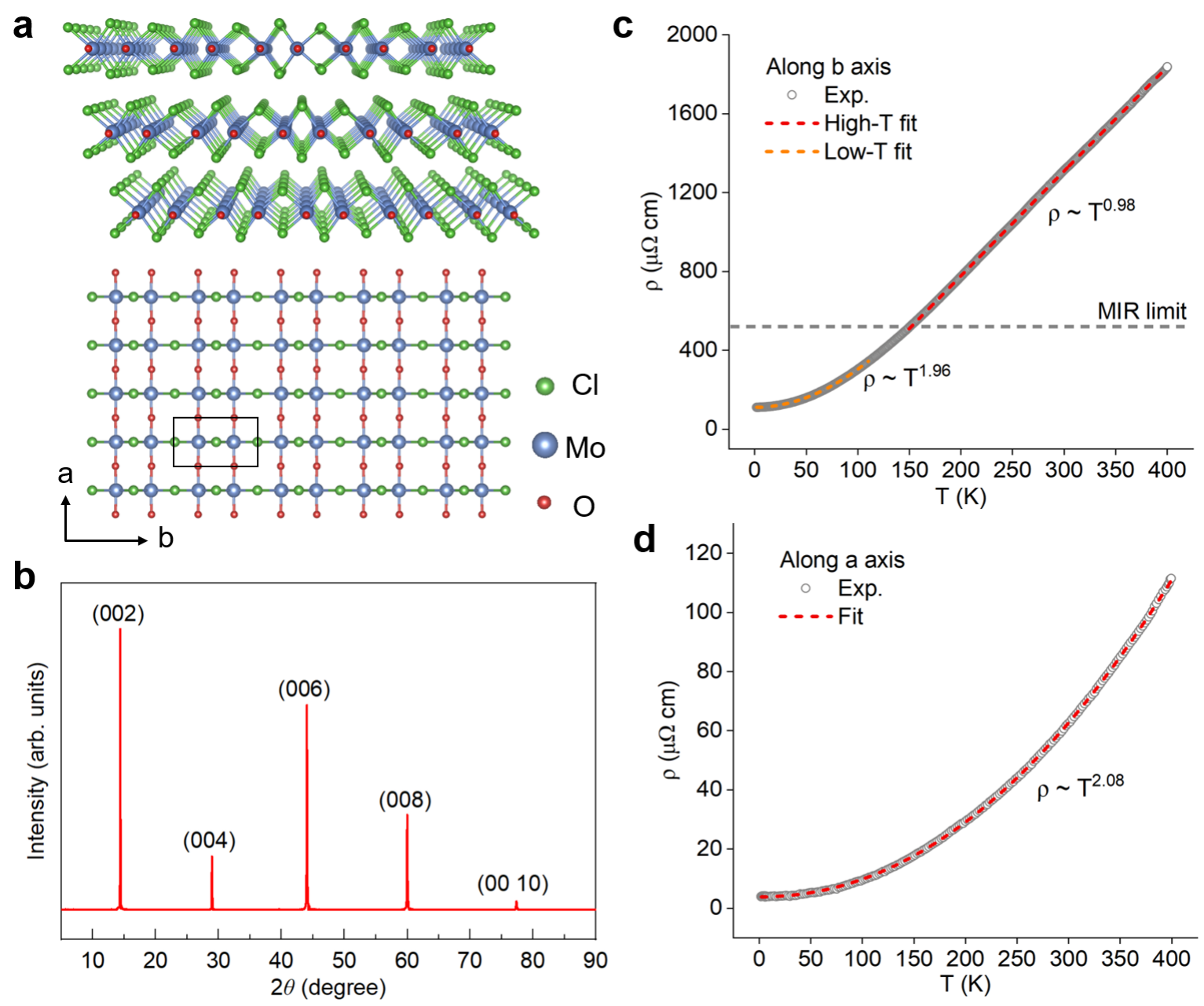

**Fig. 1 | Orthogonal electronic dichotomy in $MoOCl_2$. a**, Crystal structure of $MoOCl_2$ viewed from different crystallographic directions, showing the van der Waals layered structure and the orthogonal in-plane axes. The crystal structure is visualized using VESTA (ref.[46]). **b**, X-ray diffraction pattern of a representative crystal. **c**, Temperature-dependent resistivity along the b-axis. Fits to $\rho = \rho_0 + AT^n$ reveal a crossover from a low-temperature nearly quadratic dependence ($n = 1.96$) to an extended high-temperature nearly linear-in-$T$ regime ($n = 0.98$), characteristic of strange metal transport. The grey dashed line marks the calculated MIR resistivity limit. **d**, Temperature-dependent resistivity along the a-axis, exhibiting a nearly quadratic temperature dependence ($n = 2.08$), indicating conventional Fermi liquid transport.

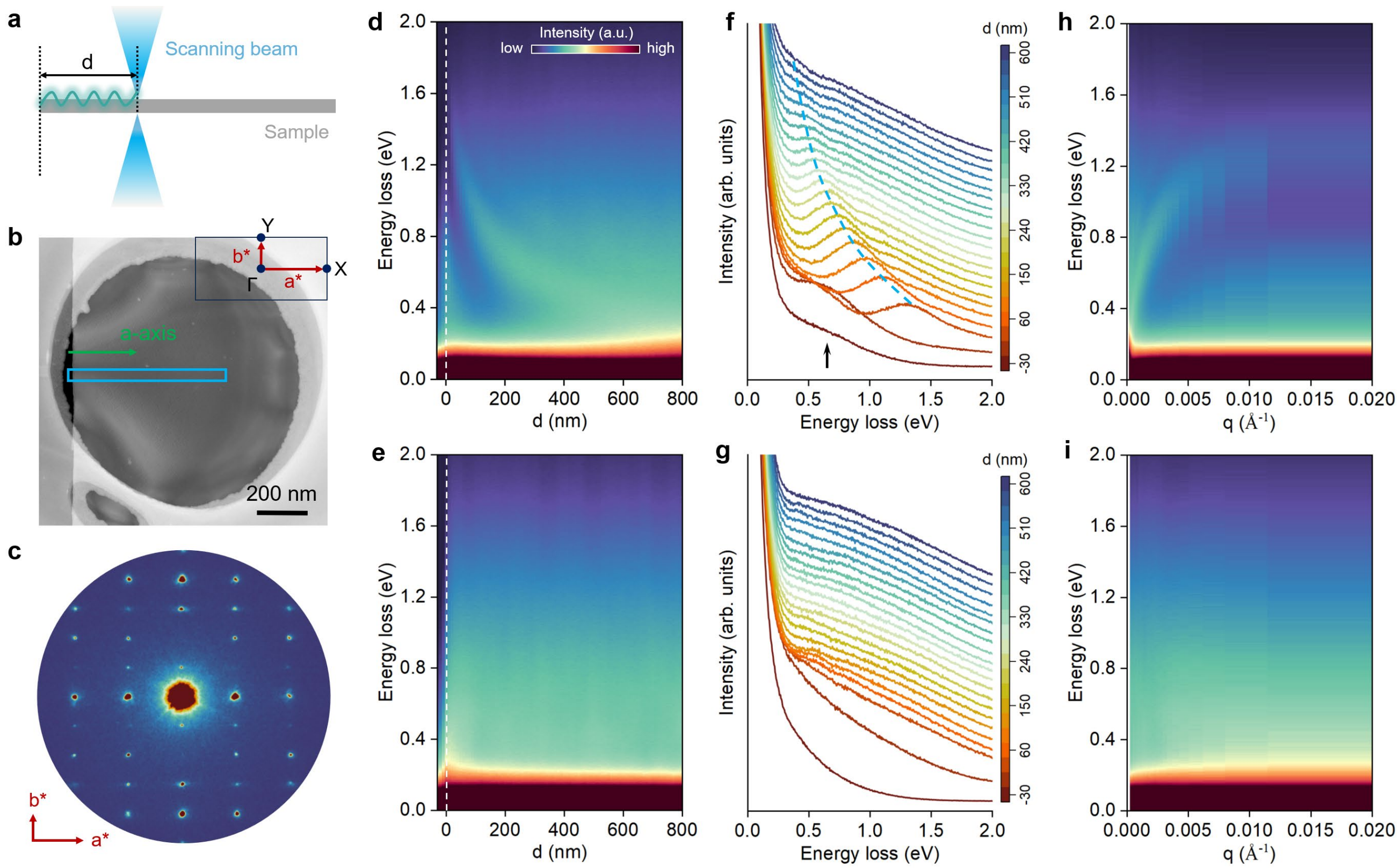

**Fig. 2 | Direction-selective propagation of plasmon polaritons. a**, Schematic of the edge-scanning geometry. The electron beam is scanned perpendicular to the sample edge at a distance $d$. **b**, Real-space image of an 8-layer $MoOCl_2$ flake showing the electron-probe scan region (marked by the light blue rectangle) near the flake edge. Inset: schematic Brillouin zone indicating the reciprocal a* and b* directions. **c**, Electron diffraction pattern used to determine the crystallographic orientation of the flake and identify the corresponding real-space a- and b-oriented edges. **d**, **e**, Energy-loss intensity maps measured as a function of distance $d$ from the edge ($d = 0$, marked by the white dashed line) along the a- and b-axis, respectively. A pronounced distance-dependent interference pattern is observed along the a-axis, indicating coherent propagating plasmon polaritons. **f**, **g**, Stacked energy distribution curves (EDCs) extracted from **d** and **e**, respectively. In **f**, the blue dashed line serves as a guide to the peak positions of the plasmon polaritons propagating along the a-axis. The black arrow marks the edge-derived polaritonic signal measured in the aloof region. **h**, **i**, Energy-momentum dispersion maps reconstructed from **d** and **e**, respectively. Only along the a-axis direction is a well-defined plasmon polariton branch observed.

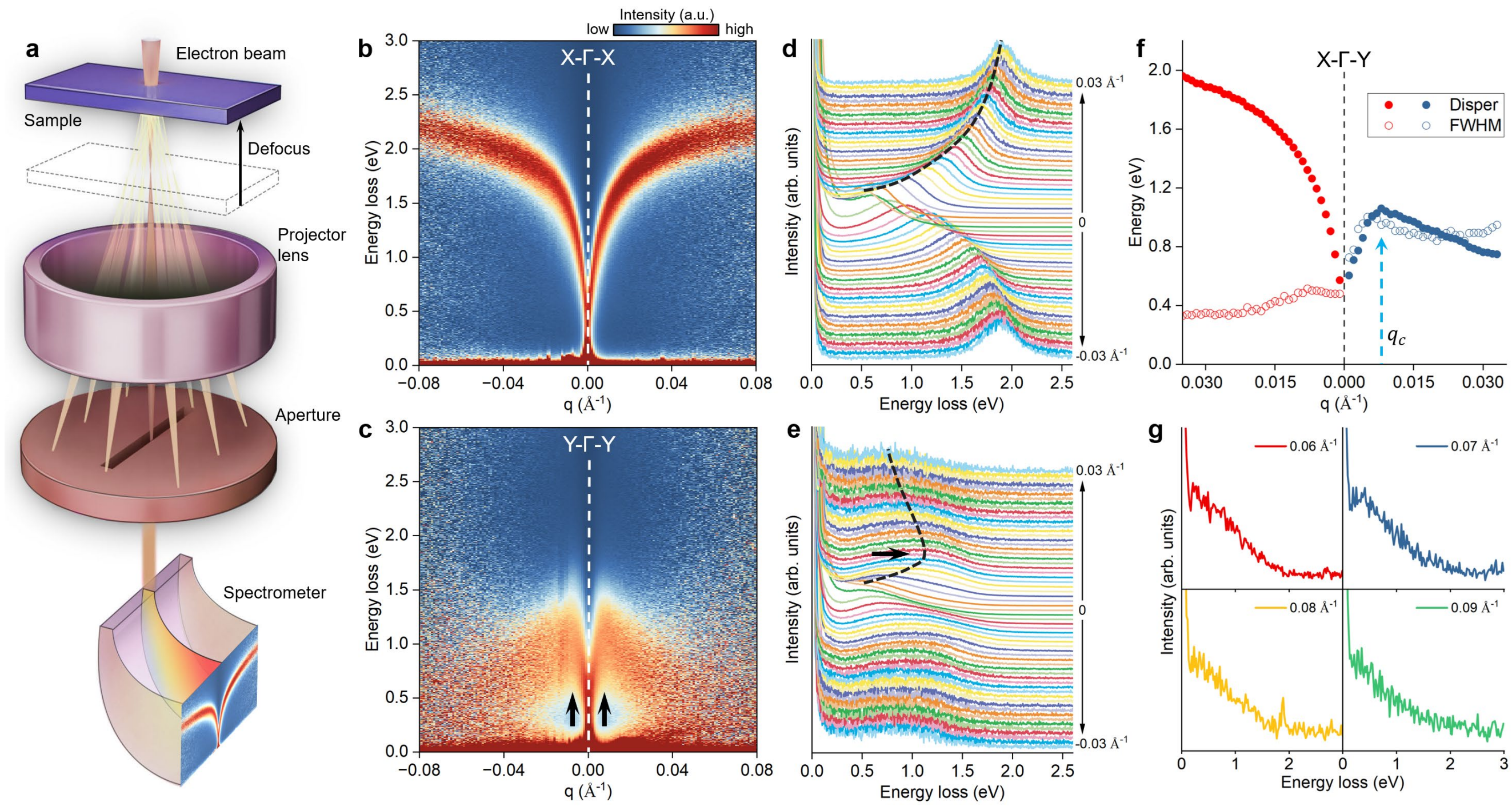

**Fig. 3 | Directional evolution and collapse of plasmons in the long-wavelength regime. a**, Schematic of the defocus-engineered configuration used to measure long-wavelength charge excitations with high momentum resolution. **b**, **c**, Energy-momentum EELS maps measured from an 8-layer $MoOCl_2$ flake along the X-Γ-X and Y-Γ-Y directions, corresponding to the Fermi liquid and strange metal directions, respectively. A well-defined plasmon branch with positive dispersion is clearly observed along the X-Γ-X direction, whereas the spectral peak rapidly broadens and loses its pole-like character along the Y-Γ-Y direction. **d**, **e**, Stacked EDCs extracted from **b** and **c**, respectively. The black dashed lines are guides to the eye. Black arrows in **c** and **e** indicate the momentum point where the dispersion bends back. **f**, Extracted dispersion and full width at half maximum (FWHM) of the charge excitations along the combined X-Γ-Y path. The critical momentum $q_c$ for dispersion bending is indicated by the blue dashed arrow. **g**, Representative high momentum EDCs along the Γ-Y direction, showing a broad, nearly featureless continuum.

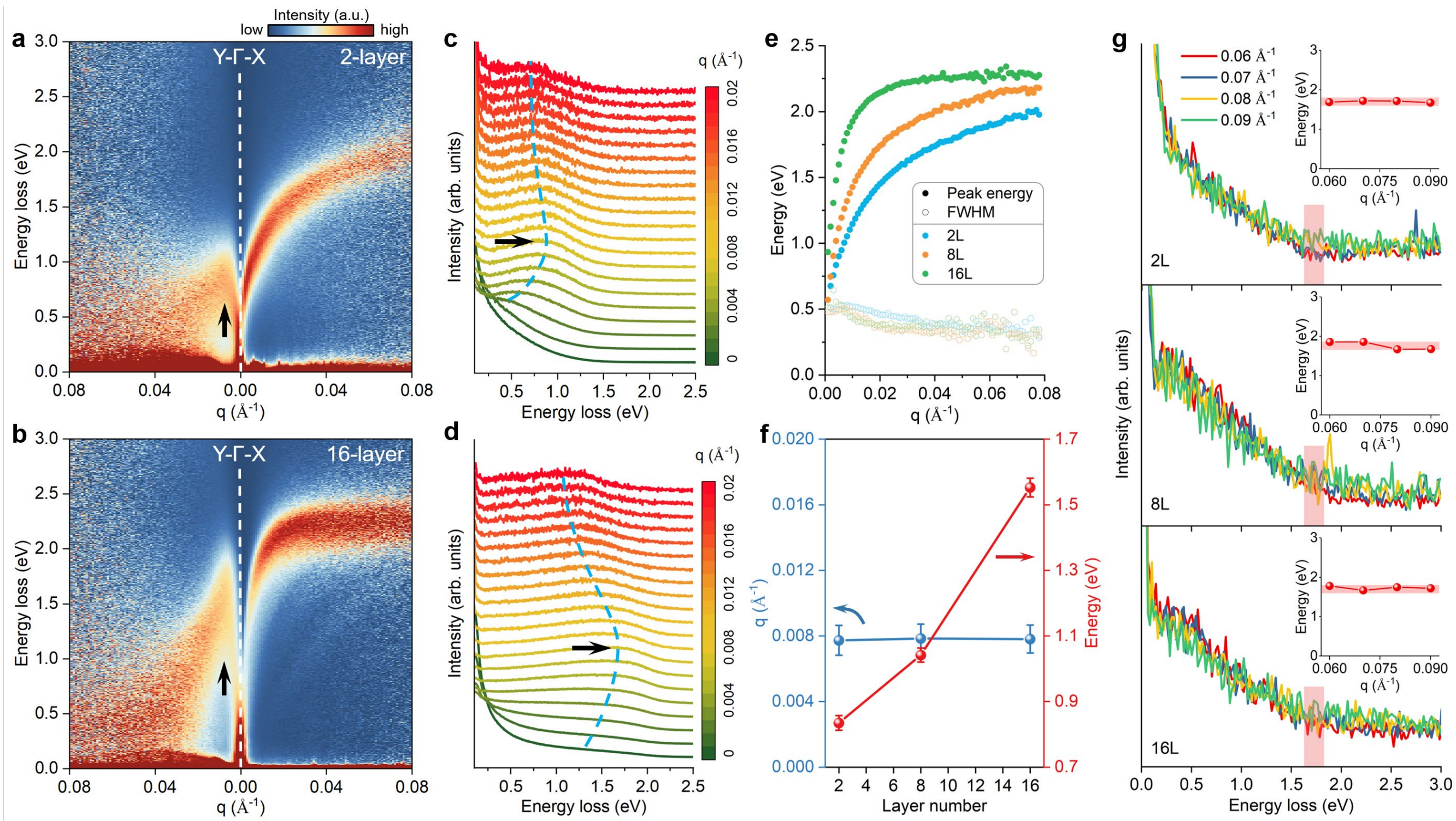

**Fig. 4 | Intrinsic scales of incoherent charge dynamics revealed by thickness-dependent measurements.** **a**, **b**, Energy-momentum EELS maps measured along the Y-Γ-X path for 2-layer and 16-layer $MoOCl_2$ flakes. **c**, **d**, Stacked EDCs along the Γ-Y direction extracted from **a** and **b**, respectively. The blue dashed lines are guides to the eye, while the black arrows mark the critical momentum point where the dispersion bends back. **e**, Extracted plasmon dispersion and FWHM along the Γ-X direction for 2-layer, 8-layer and 16-layer $MoOCl_2$. **f**, Thickness dependence of $q_c$ (blue spheres, left axis) and the peak energy of the charge excitations at $q_c$ (red spheres, right axis) along the Γ-Y direction. $q_c$ remains nearly unchanged with increasing layer number, whereas the corresponding excitation energy increases significantly. The error bars denote the FWHM of the elastic peak. **g**, Representative high momentum EDCs acquired along the Γ-Y direction for the three thicknesses. The shaded region marks the high-energy boundary $E_c$. The insets show the extracted $E_c$ as a function of $q$ for different thicknesses. Its negligible variation with either momentum or thickness identifies $E_c$ as an intrinsic energy scale characterizing the incoherent charge dynamics.

## Data availability

The data that support the findings of this study are available from the corresponding authors on request.

## Acknowledgements

We thank L. Xian, S. Xue, and X. Liu for their discussions. P.G. acknowledges the support from the New Cornerstone Science Foundation through the XPLORER PRIZE.

## Author contributions

P.G. supervised the project. J.L. conceived the project. D.H. grew the single-crystal sample with the guidance of Z.W.W.; J.W. prepared the TEM sample. J.L. and P.H. designed and performed the STEM-EELS measurements. D.H. and J.L. designed and performed the transport measurements with the help of J.S. and W.S.D.; J.L. carried out the theoretical calculations in consultation with X.Z. and J.G.; J.L. performed the data processing and analysis. J.L. wrote the paper with input from P.H. and P.G.

## Funding

This work was supported by the National Natural Science Foundation of China (Grant Nos. 52125307, 12504198), the National Key Research and Development Program of China (Grant Nos. 2025YFA1411200, 2022YFA1403400, 2021YFA1400503), the Beijing National Laboratory for Condensed Matter Physics (Grant No. 2023BNLCMPKF007), the Zhenjiang Science&Technology Program (Grant No. JC 2024003), and the Guangdong Basic and Applied Basic Research Foundation (Grant No. 2024A1515030118).

## Competing interests

The authors declare no competing interests.